\documentclass[apj]{emulateapj}
\usepackage{amsmath}
\usepackage{amsfonts}
\usepackage{amsthm}
\usepackage{amssymb}
\usepackage{graphicx}
\usepackage{mathrsfs}
\usepackage{latexsym}
\usepackage{graphics}
\usepackage{cases}
\usepackage{hyperref}
\usepackage{mathtools}
\usepackage{dsfont}
\usepackage[percent]{overpic}

\usepackage{color}
\usepackage{ulem}

\newcommand{\be}{\begin{equation}}
\newcommand{\ee}{\end{equation}}
\newcommand{\bea}{\begin{eqnarray}}
\newcommand{\eea}{\end{eqnarray}}
\newcommand{\bes}{\begin{equation}\begin{split}}
\newcommand{\ees}{{\end{split}\end{equation}}}
\renewcommand{\d}{d}
\newcommand{\logm}{\tilde{m}}
\newcommand{\logv}{\tilde{v}}
\newcommand{\logw}{\tilde{w}}
\newcommand{\logs}{\tilde{s}}
\newcommand{\tfr}{\mathbf{T}}

\newcommand{\ha}{H{\sc\,i}}

\newcommand{\erf}{\rm erf}

\newcommand{\kms}{{\rm km~s^{-1}}}
\newcommand{\msun}{{\rm M}_{\odot}}

\newcommand{\ensavg}[1]{\langle#1\rangle}
\newcommand{\avg}[1]{\overline{#1}}
\newcommand{\abs}[1]{|#1|}
\newcommand{\eq}[1]{Equation~(\ref{eq_#1})}
\newcommand{\fig}[1]{Figure~\ref{fig_#1}}
\newcommand{\tab}[1]{Table~\ref{tab_#1}}
\newcommand{\col}{\\[1ex](A color version of this figure is available in the online journal.)}

\begin{document}

\title{PRECISE TULLY-FISHER RELATIONS WITHOUT GALAXY INCLINATIONS}

\author{D. Obreschkow$^1$}
\author{M. Meyer$^1$}
\affiliation{$^1$International Centre for Radio Astronomy Research (ICRAR), M468, University of Western Australia, 35 Stirling Hwy, Crawley, WA 6009, Australia}


\date{\today}

\begin{abstract}
Power-law relations between tracers of baryonic mass and rotational velocities of disk galaxies, so-called Tully-Fisher relations (TFRs), offer a wealth of applications in galaxy evolution and cosmology. However, measurements of rotational velocities require galaxy inclinations, which are difficult to measure, thus limiting the range of TFR studies. This work introduces a maximum likelihood estimation (MLE) method for recovering the TFR in galaxy samples with limited or no information on inclinations. The robustness and accuracy of this method is demonstrated using virtual and real galaxy samples. Intriguingly, the MLE reliably recovers the TFR of all test samples, even without using any inclination measurements -- that is, assuming a random $\sin i$-distribution for galaxy inclinations. Explicitly, this `inclination-free MLE' recovers the three TFR parameters (zero-point, slope, scatter) with statistical errors only about 1.5-times larger than the best estimates based on perfectly known galaxy inclinations with zero uncertainty. Thus, given realistic uncertainties, the inclination-free MLE is highly competitive. If inclination measurements have mean errors larger than $10^\circ$, it is better not to use any inclinations, than to consider the inclination measurements to be exact. The inclination-free MLE opens interesting perspectives for future \ha\ surveys by the SKA and its pathfinders.
\end{abstract}


\maketitle


\section{Introduction}\label{section_introduction}

An astronomical scaling relation of great historical merit is the Tully-Fisher relation \citep[TFR,][]{Tully1977}, an empirical scaling law between the absolute magnitude of disk galaxies and their edge-on spectral linewidths. This relation originates from a fundamental connection between baryon mass, roughly traced by luminosity, and circular velocity, imprinted in the Doppler-broadening of emission lines. Today, any power-law relation between a mass tracer and a proxy for the circular velocity of disk galaxies is called a TFR, such as the optical/infrared TFR \citep{Pizagno2007,Meyer2008}, the stellar mass TFR \citep{Kassin2007}, the neutral atomic hydrogen (\ha) TFR \citep{Chakraborti2011}, and the baryonic TFR \citep{McGaugh2000}. Such TFRs can be used to constrain the dark matter potential of galaxies \citep{Trachternach2009}, their gas-content \citep{McGaugh1997,Pfenniger2005}, their luminosity evolution \citep{Ziegler2002,Boehm2004}, the stellar initial mass function \citep{Bell2001}, the formation of central bars \citep{Courteau2003}, galactic outflows \citep{Dutton2012}, the transport of angular momentum between halos and disks \citep{Sommer-Larsen2001}, standard dark matter models \citep{Obreschkow2013b}, and alternative theories of gravity \citep{Chakraborti2011,McGaugh2012}. Adopting the TFR as a prior, observed \ha\ linewidths can be converted into absolute magnitudes, which, given apparent magnitudes, provide redshift-independent distances. Combined with spectroscopic redshifts ($z$), these can constrain the cosmic expansion rate \citep{Tully1977} and superposed peculiar motions \citep{Masters2006}, including the local bulk flow \citep{Nusser2011}.

Circular velocities of galaxies are normally derived from the width of the total 21~cm emission line of \ha, because most \ha\ lies beyond the stellar scale radius, hence better tracing the asymptotic circular velocity than optical emission lines. The apparent \ha\ linewidth must then be corrected for the orbital inclination $i\in[0,\pi/2]$, defined as the angle between the line-of-sight and the rotational axis of the disk. This inclination is inferred from optical images, because the \ha\ detections tend to be spatially unresolved. Even future interferometric \ha\ surveys with the Square Kilometer Array (SKA) and its pathfinders will only marginally resolve most \ha\ detections \citep[e.g.,][for ASKAP]{Duffy2012}. However, the need for optical inclinations leads to serious issues.

First, optical inclinations may be simply unavailable. For example, the optical counterparts of the \ha\ Parkes All Sky Survey \citep[HIPASS,][]{Barnes2001} have insufficient sensitivity to recover the inclinations of most disks with circular velocities below $50~\kms$ \citep{Obreschkow2013b} and dust extinction and stellar confusion in the galactic plane cause poor completeness \citep{Doyle2005}. Future high-$z$ \ha\ surveys with the SKA will need parallel sub-arcsecond imaging to even barely resolve typical $L^*$-galaxies (at $z\approx1$). The hence required active optics-assisted or space-based observations might yield insufficient sky coverage.

Second, optical inclinations are uncertain; they often represent the largest uncertainty in recovering circular velocities. Unless kinematic maps are available, inclinations are mostly obtained by fitting ellipses to the isophotes, giving an apparent axis ratio $q$. Upon assuming a spheroidal disk with an intrinsic axis ratio $q_0$, $i$ is then given by 
\be\label{eq_cosi}
	\cos^2i=\frac{q^2-q_0^2}{1-q_0^2}\,.
\ee
The measurement of $q$ is deteriorated by smearing effects \citep{Giovanelli1997b,Masters2003} and physical features such as disk asymmetries and differential dust absorption. The value of $q_0$ can only be approximated from local edge-on galaxies and its dependence on the Hubble type remains debated \citep{Tully2009}. At high $z$ the determination of $q_0$ is further complicated by the cosmic evolution of the disk thickness \citep{Bournaud2007} and the bulge-to-disk ratio \citep{Driver2013}.

Third, optical inclinations can differ significantly from \ha\ inclinations. Examples are galaxies with warped \ha\ disks \citep[e.g., NGC 4013,][]{Bottema1987}, which are more the rule than the exception \citep{Sancisi1976}. 

The objective of this paper is to establish a new method to derive TFRs using limited or no information on inclinations. This method can recover the most likely TFR, assumed to be a power-law with Gaussian scatter, given a galaxy sample where the inclination of each object $k$ is described by a probability density function (PDF) $\rho_k(i)$. This method can then be applied to several cases: (1) a galaxy sample with approximate inclination measurements for every galaxy; (2) a galaxy sample without individual inclination measurements, but with a known prior PDF $\rho(i)$, obtained, for example, from deep imaging of a small subsample; and (3) a galaxy sample without any prior information on inclinations. In the latter case, the galaxy axes can be assumed isotropically oriented, that is $\rho(i)=\sin(i)$.

This paper first derives a generic maximum likelihood estimation (MLE) to obtain a TFR using limited or no knowledge on galaxy inclinations (Section \ref{section_method}). This new method is then tested and characterized in detail (Section \ref{section_simulation}) using idealized control samples of virtual galaxies. In particular, these tests allow a comparison of the classical approach, where slightly erroneous inclinations are treated as exact, against a version of the MLE that does not use any inclinations at all. In Section \ref{section_observation}, we then apply this new MLE approach to the local $K$-band and $B$-band TFRs studied in detail by \cite{Meyer2008}. The key findings are summarized in Section \ref{section_conclusion}.

\section{Analytical method}\label{section_method}

\subsection{Derivation of the maximum likelihood approach}\label{subsection_generic_mle}

This section introduces our generic method to estimate TFRs using uncertain/unknown inclinations. Here, we focus on the basic model of a linear TFR with Gaussian scatter (in log-space). Extensions to this model, e.g., to deal with non-linear TFRs and non-Gaussian scatter, are described in Appendix \ref{appendix_extensions}.

The basic TFR can be written as
\be\label{eq_observable_tfr}
	\logm = \alpha+\beta\logv,
\ee
where tildes denote logarithms, i.e.,~$\tilde{x}\equiv\log_{10}x$, $m$ is a tracer of the galaxy mass, e.g., an optical luminosity, a gas mass, the stellar mass, or the baryon mass, and $v$ is a tracer of the circular velocity. The parameters $\alpha$ (zero-point) and $\beta$ (slope) are scalar constants. The `true' values $\logm_k$ and $\logv_k$ of real galaxies scatter around \eq{observable_tfr} with an ensemble variance $\sigma^2\equiv\ensavg{(\alpha+\beta\logv_k-\logm_k)^2}$. Its square-root $\sigma$ is called the intrinsic scatter. Upon assuming that this scatter derives from a log-normal distribution, the conditional PDF per unit $\logv$ for a galaxy with mass $m$ to exhibit a circular velocity $v$ is given by
\be\label{eq_rho_tfr}
	\rho(\logv|\logm,\tfr) = \frac{\beta}{\sqrt{2\pi}\sigma}\exp\Big[-\tfrac{(\logm-\alpha-\beta\logv)^2}{2\sigma^2}\Big].
\ee
In this expression and for the rest of this work
\be\label{eq_def_tfr}
	\tfr\equiv(\alpha,\beta,\sigma)
\ee
is the vector grouping the three TFR parameters.

The circular velocity $v$ of a rotating disk cannot be measured directly, since the Doppler-effect only traces the line-of-sight projection
\be\label{eq_wv}
	w=v\sin i,
\ee
where $i$ is the orbital inclination of the material that $v$ refers to. In a simplistic galaxy model with a \ha\ disk of constant circular velocity $v$ and zero dispersion, $w$ is given by the width $W$ of the \ha\ emission line via $w=W/2$. When dealing with real \ha\ emission lines with a dispersive component, $W$ is often substituted for $W_{20}$ \citep{McGaugh2000} or $W_{50}$ \citep{Kannappan2002}, where $W_p$ is the \ha\ linewidth measured at $p$ percent of the peak flux density. More elaborate equations can also account for cosmological and instrumental line broadening, as well as remove turbulent motion from the linewidth \citep[e.g.][]{Tully1985,Verheijen2001b,Springob2007,Meyer2008}. In this section, there is no need to specify the definition of $w$. We only assume that $w$ is defined and measured in some way and used with \eq{wv} to compute $v$.

Given a sample of $N$ galaxies $k=1,...,N$ and upon assuming no prior information on $\tfr$, the most likely $\tfr$ maximizes the global likelihood function $\mathcal{L}(\tfr)$, conveniently written in log-form,
\be\label{eq_likelihood}
	\ln\mathcal{L}(\tfr)=\sum_{k=1}^N g_k \ln\rho_k(\tfr),
\ee
where $\rho_k(\tfr)$ is the conditional PDF of observing a galaxy with the properties of galaxy $k$ given the Tully-Fisher parameters $\tfr$. The scalars $g_k$ are normalized weights, such that $\sum g_k=1$. If all galaxies are considered equally important, then $g_k=N^{-1}~\forall k$. To estimate the TFR of a sample that is complete inside a galaxy-dependent volume $V_k$, the weights could be chosen as $g_k\propto V_k^{-1}$.

If each galaxy had perfectly measured values $\{\logm_k,\logw_k\}$, $\rho_k(\tfr)$ would be equal to the conditional PDF $\rho_k(\logw_k,\logm_k|\tfr)$ of finding $\{\logm_k,\logw_k\}$ given the TFR with parameters $\tfr$. However, in practice, the true masses and projected velocities differ from the measured values $\logm_k$ and $\logw_k$ by measurement errors. The true values can only be described in terms of finite PDFs $\rho_k(\logm)$ and $\rho_k(\logw)$, respectively. In that case, $\rho_k(\tfr)$ can generally be written as
\be\label{eq_rhok_general}
	\rho_k(\tfr) = \iint d\logm\,d\logw\,\rho_k(\logm)\,\rho_k(\logw)\,\rho_k(\logw,\logm|\tfr).
\ee

The remaining task is to express \eq{rhok_general} in a usable analytical form. The rule of joint probabilities implies $\rho_k(\logm,\logw|\tfr)=\rho_k(\logw|\logm,\tfr)\rho_k(\logm|\tfr)$. Here, we assume no prior constraints on $\logm$, thus $\rho_k(\logm|\tfr)$ can be taken as unity; hence, $\rho_k(\logm,\logw|\tfr)=\rho_k(\logw|\logm,\tfr)$. Summing over all possible inclinations $i$ then implies $\rho_k(\logw|\logm,\tfr)=\int_0^{\pi/2}\d i\,\rho_k(i)\,\rho_k(\logv|\logm,\tfr)$, where $v=w\,(\sin i)^{-1}$ and $\rho_k(i)$ is the normalized prior PDF of the inclination $i_k\in[0,\pi/2]$.
Using Equations~(\ref{eq_rho_tfr}) and (\ref{eq_wv}),
\be\label{eq_main_rho1}
\begin{split}
	 \rho_k(\logw,\logm|\tfr)=\,&\frac{\beta}{\sqrt{2\pi}\sigma}\int_0^{\pi/2}\!\!\!\!\d i\,\rho_k(i) \times \\
	&\exp\Big[-\tfrac{(\logm-\alpha-\beta\logw+\beta\log_{10}\sin i)^2}{2\sigma^2}\Big].
\end{split}	
\ee
This PDF is indexed with $k$, since in general different galaxies may have different inclination priors $\rho_k(i)$. 

Upon assuming that the measurement uncertainties of $\logm_k$ and $\logw_k$ are Gaussian with galaxy-dependent standard deviations $\sigma_{\logm_k}$ and $\sigma_{\logw_k}$, \eq{rhok_general} can be solved\footnote{Using $\rho_k(\logm)=\exp[-(\logm-\logm_k)^2/(2\sigma_{\logm_k}^2)]/(\sqrt{2\pi}\sigma_{\logm_k})$ and $\rho_k(\logw)=\exp[-(\logw-\logw_k)^2/(2\sigma_{\logw_k}^2)]/(\sqrt{2\pi}\sigma_{\logw_k})$ in \eq{rhok_general} and applying the convolution theorem results in \eq{main_rho2}.},
\be\label{eq_main_rho2}
\begin{split}
	\rho_k(\tfr) =\,&\frac{\beta}{\sqrt{2\pi}\sigma_{tot,k}}\int_0^{\pi/2}\!\!\!\!\d i\,\rho_k(i) \times \\
	&\exp\Big[-\tfrac{(\logm_k-\alpha-\beta\logw_k+\beta\log_{10}\sin i)^2}{2\sigma_{tot,k}^2}\Big],
\end{split}	
\ee
where the total scatter $\sigma_{tot,k}\equiv(\sigma^2+\sigma_{\logm_k}^2+\beta^2\sigma_{\logw_k}^2)^{1/2}$ is the combination of the intrinsic scatter $\sigma$ of the TFR and the observational scatter $(\sigma_{\logm_k}^2+\beta^2\sigma_{\logw_k}^2)^{1/2}$.

In summary, if every galaxy $k$ has measured values $\logm_k$ and $\logw_k$ with Gaussian uncertainties $\sigma_{\logm_k}$ and $\sigma_{\logw_k}$, as well as an inclination $i_k$ specified by the PDF $\rho_k(i)$, then the most likely TFR is found by maximizing \eq{likelihood} with $\rho_k(\tfr)$ given in \eq{main_rho2}. If the measurement uncertainties are highly non-Gaussian, then the full integral of \eq{rhok_general} should be used instead of \eq{main_rho2}. If all $\logm_k$ and $\logw_k$ are measured perfectly ($\sigma_{\logm_k}=\sigma_{\logw_k}=0$), \eq{main_rho2} remains valid and $\sigma_{tot,k}$ reduces to the intrinsic scatter $\sigma$.

\subsection{Practical considerations}\label{subsection_real_data}

The formalism of \S\ref{subsection_generic_mle} assumes an idealized scenario with perfect data bar Gaussian measurement uncertainties. This section explains how to deal with two typical deviations from this scenario, namely (i) inclination-dependent extinction and (ii) non-Gaussian outliers in the data. More advanced extensions, dealing with complex detection limits, early-type galaxies, and non-linear TFRs, are sketched out in Appendix \ref{appendix_extensions}.

\textit{Inclination-dependent extinction}: The internal extinction of galaxies depends on their inclination $i$ \citep[e.g., ][]{Giovanelli1994}, meaning that the derived mass $\logm$ depends on the inclination $i$. Sometimes this extinction can be neglected -- typically for the \ha\ line and rest-frame near-infrared continuum \citep{Gordon2003}. In most cases, however, extinction is significant. If extinction is important and the inclination $i_k$ is not precisely known, then $\logm_k$ is not a well-defined value, but of a function $\logm_k(i)$,~$i\in[0,\pi/2]$, for every galaxy. In that case, it suffices to substitute $\logm_k$ in \eq{main_rho2} for $\logm_k(i)$. For example, let's consider $\logm_k$ to be a mass computed from the absolute magnitude $M_{k,\lambda}(i)$ at wavelength $\lambda$. This magnitude is estimated from the observed absolute magnitude $M_{k,\lambda}$ via $M_{k,\lambda}(i)=M_{k,\lambda}-\Delta M_{\lambda}(i)$, where $\Delta M_{\lambda}(i)$ is the extinction factor. Simple empirical models for $\Delta M_{\lambda}(i)$ in the $B$, $R$, $J$, $H$, and $K$ bands are summarized in \S3.1.4 of \cite{Meyer2008} \citep[based on][]{Tully1998,Masters2003}, and extended models were made available by \cite{Driver2008}, \cite{Maller2009}, and \cite{Masters2010}. Using those models, observed absolute magnitudes can be translated into $\logm_k(i)\propto M_{k,\lambda}-\Delta M_{\lambda}(i)$. This approach will be used in Section \ref{section_observation} when recovering the $K$-band and $B$-band TFR without using inclination measurements.

\textit{Non-Gaussian outliers}: Data with erroneous outliers, such as misidentifications between \ha\ sources (used to measure $w$) and optical counterparts (used to measure $m$), can adversely affect the estimation of a TFR. Of course, this applies to any method used to estimate a TFR, not just to the MLE introduced in this work. However, when dealing with galaxies of uncertain/unknown inclinations $i_k$ -- a particular strength of the MLE -- outliers are hard to spot. This is because without inclinations measurements the data can only be shown in the $(\logw,\logm)$-plane but not in the $(\logv,\logm)$-plane. If the dataset is likely to include some outliers, the likelihood function (\eq{likelihood}) should be maximized using a robust technique, which ignores the $fN$ galaxies with the lowest likelihood terms, where $f<1$ is the estimated upper bound for the fraction of outliers.

\subsection{Special case: completely unknown inclinations}

In the case where galaxy inclinations are completely unknown, the MLE of \S\ref{subsection_generic_mle} can be used to recover the most likely TFR by choosing $\rho_k(i)=\sin i$. The $\sin i$-distribution is that expected from an isotropic random orientation of galaxies, hence encoding our complete lack of knowledge on individual galaxy orientations. Hereafter, this method will be called the `inclination-free' maximum-likelihood estimation (MLE). If a better prior for $\rho_k(i)$ is available, for example from inclination measurements in a subsample of galaxies or from modelling the survey properties, then this prior can be used instead (see quantitative analysis in \S\ref{subsection_inclination_distribution_errors}). 

\subsection{Special case: perfectly known inclinations}\label{subsection_MLE_with_inclinations}

Let us now investigate the opposite extreme, where all galaxy inclinations $i_k$ are precisely known -- the default assumption in most current literature on TFRs. In this case, the extinction-corrected masses $\logm_k$ and deprojected velocities $\logv_k=\logw_k-\log_{10}\sin i_k$ are known up to the observational uncertainties $\sigma_{\logm_k}$ and $\sigma_{\logv_k}=\sigma_{\logw_k}$.

It is an occasional misconception that the TFR can be accurately recovered by fitting a linear regression to the data in the $(\logv,\logm)$-plane. To show that regressions are not generally appropriate one can depart from \eq{main_rho2} assuming that the inclinations $i_k$ are precisely known; thus $\rho_k(i)=\delta(i_k-i)$, where $\delta$ is the Dirac-delta function. \eq{main_rho2} then reduces to
\be\label{eq_rho_tfr2}
	\rho_k(\tfr) = \frac{\beta}{\sqrt{2\pi}\sigma_{tot,k}}\exp\Big[-\tfrac{(\logm_k-\alpha-\beta\logv_k)^2}{2\sigma_{tot,k}^2}\Big].
\ee
Hence the likelihood function $\mathcal{L}(\tfr)$ of \eq{likelihood} is maximized by minimizing
\be\label{eq_likelihood_fixedi}
	\sum_{k=1}^N g_k\left[\ln\frac{\sigma_{tot,k}}{\beta}+\frac{(\logm_k-\alpha-\beta\logv_k)^2}{2\sigma_{tot,k}^2}\right].
\ee

Only upon neglecting the contribution of $\ln(\sigma_{tot,k}/\beta)$ and assuming vanishing intrinsic scatter, $\sigma\rightarrow0$, \eq{likelihood_fixedi} is minimized by minimizing
\be\label{eq_likelihood_fixedi_simple}
	\chi^2\equiv\sum_{k=1}^N g_k \frac{(\logm_k-\alpha-\beta\logv_k)^2}{\sigma_{\logm_k}^2+\beta^2\sigma_{\logv_k}^2}\,.
\ee
This minimization corresponds to a bivariate linear regression. It reduces to a standard linear regression, if all $\sigma_{\logm_k}$ are identical and all $\sigma_{\logv_k}=0$.

The difference between Equations (\ref{eq_likelihood_fixedi}) and (\ref{eq_likelihood_fixedi_simple}) shows that linear regressions, even bivariate ones, do not generally recover the most likely TFR. Instead, \eq{likelihood_fixedi} must be minimized by simultaneously varying all three parameters $\alpha$, $\beta$, and $\sigma$. In practice, the errors made by using regression techniques are often small -- perhaps smaller than the statistical uncertainties of the Tully-Fisher parameters $\tfr$. However, as TFR measurements become increasingly accurate, it seems worth noting that linear regressions techniques are not optimal (illustrations in \S\ref{subsection_convergence}).

\begin{figure}[t]
	\includegraphics[trim=31 10 47 30,clip,width=\columnwidth]{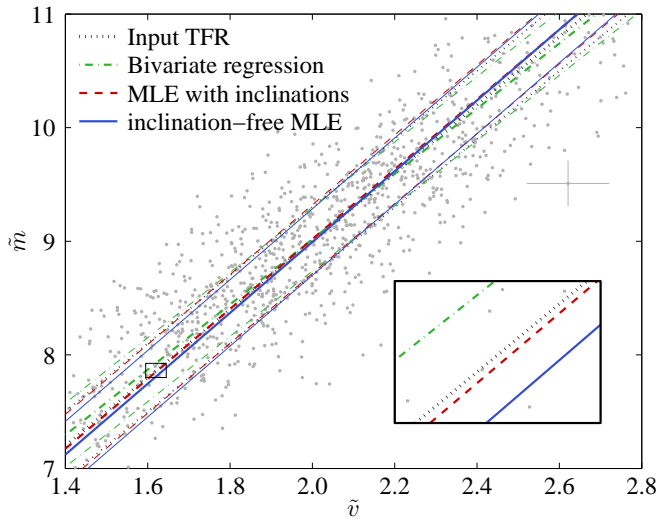}
	\caption{Illustration of a galaxy control sample (points) drawn from an input TFR shown as black dotted lines. The TFR has an intrinsic vertical scatter of 0.3 dex. In addition, each data point has been randomly perturbed by a random measurement error in both coordinates, represented by the 68\%-error bars shown for one point. The colored lines show the TFRs recovered using the different methods detailed in \S\ref{subsection_basic_example}. For each line type there are three lines representing the mean TFR ($\logm=\alpha+\beta\logv$) and the $1\sigma$-scatter intervals ($\logm=\alpha+\beta\logv\pm\sigma$). This figure only provides a qualitative illustration of the different methods, since the recovered TFRs depend on the random realization of the sample. A quantitative assessment of the different methods is given in \fig{errors}.\col}
	\label{fig_basic_example}
\end{figure}

\section{Verification using mock data}\label{section_simulation}

This section tests the accuracy and liability of the MLE to recover the TFR using idealized control samples of model-galaxies.

\subsection{Control sample}\label{subsection_control_sample}

The control samples consist of $N$ model-galaxies $k=1,...,N$ with identical weights $g_k=N^{-1}$ and random `observed' log-masses $\logm_k$ drawn from a normal distribution with mean $9$ and unit variance. Log-normal mass distributions roughly describe the empirical mass distributions in sensitivity limited surveys (e.g., absolute magnitude distributions in SDSS, see Figure 5 in \citealp{Montero-Dorta2009}; \ha\ mass distribution in HIPASS, see Figure 1 bottom in \citealp{Zwaan2003}). The low-mass drop-off is generally due to the sensitivity limit, thus small effective volume, while the high-mass drop-off is caused by the steepness of the galaxy mass function.

The `true' log-masses $\logm_k^t=\logm_k-\Delta \logm_k$ differ from the observed log-masses $\logm_k$ by a measurement error $\Delta \logm_k$, drawn from a normal distribution with $\sigma_{\logm}=0.2$. Given the true masses $m_k^t$, the corresponding circular velocities $v_k^t$ are drawn randomly from a TFR, i.e.~from the PDF of \eq{rho_tfr}, with input values $\tfr_0=(\alpha_0,\beta_0,\sigma_0)=(3,3,0.3)$. These values roughly coincide with those of the baryonic TFR, if $m_k$ and $v_k$ are given units of $\msun$ and $\kms$ \citep{Gurovich2010}, but the precise choice of $\tfr_0$ does not matter here. Every galaxy is also assigned a random inclination $i_k$, drawn from a $\sin i$-distribution, unless otherwise specified. The $\sin i$-distribution is expected from galaxies having random orientations in three dimensions. The `true' line-of-sight projected velocities $w_k^t$ are then computed as $w_k^t=v_k^t \sin i_k$. Finally, measurement errors $\Delta \logw_k$ are drawn from a normal distribution with $\sigma_{\logw}=0.1$ to compute the observables $\logw_k=\logw_k^t+\Delta\logw_k$  and $\logv_k=\logv_k^t+\Delta\logw_k$. Inclination-dependent extinction is neglected here but will be considered in Section \ref{section_observation}.

\subsection{Basic example}\label{subsection_basic_example}

To \textit{qualitatively} illustrate the different methods to estimate a TFR, we use a single realization of a control sample (\S\ref{subsection_control_sample}) with $N=10^3$ galaxies, displayed in \fig{basic_example}. The four line-types represent four different TFRs and their 1$\sigma$-scatter: ($\cdots$) the input TFR with parameters $\tfr_0$ used to construct the data points; ($\cdot-$) the TFR recovered by fitting a bivariate linear regression to the $(\logv_k,\logm_k)$-pairs, i.e., $\alpha$ and $\beta$ minimize \eq{likelihood_fixedi_simple} and $\sigma^2=\avg{(\alpha+\beta\logv_k-\logm_k)^2}-\sigma_{\logm}^2-\beta^2\sigma_{\logv}^2$, where the over-line represents the average over all galaxies; ($--$) the TFR recovered by applying the MLE to all $(\logv_k,\logm_k)$-pairs, i.e., this TFR minimizes \eq{likelihood_fixedi}; and (---) the TFR recovered by applying the inclination-free MLE to \textit{all} $(\logw_k,\logm_k)$-pairs, i.e., this TFR maximizes \eq{likelihood}, where $\rho_k(\tfr)$ is given in \eq{main_rho2} and $\rho_k(i)=\sin i$ expresses our lack of information on the inclinations $i_k$. We emphasize that the methods ($\cdot-$) and ($--$) implicitly use the inclinations $i_k$, since in practice the $v_k$-values are recovered from $w_k$ and $i_k$ via \eq{wv}. In turn, the inclination-free MLE (---) truly disregards the galaxy inclinations by exclusively relying on the $(\logw_k,\logm_k)$-pairs.

Qualitatively, all three methods recover the input TFR fairly well. Surprisingly, even the MLE (---) that only uses the projected velocities $\logw_k$ while ignoring the inclinations $i_k$ closely recovers the input TFR. This method even outperforms the bivariate linear regression. As expected, the MLE ($--$), which uses the deprojected velocities $\logv_k$, performs better than the inclination-free MLE (---). In the following, the performance of these methods will be properly quantified.

\begin{figure*}[t]
	\centering
	\begin{tabular}{cc}
		\begin{overpic}[width=\columnwidth]{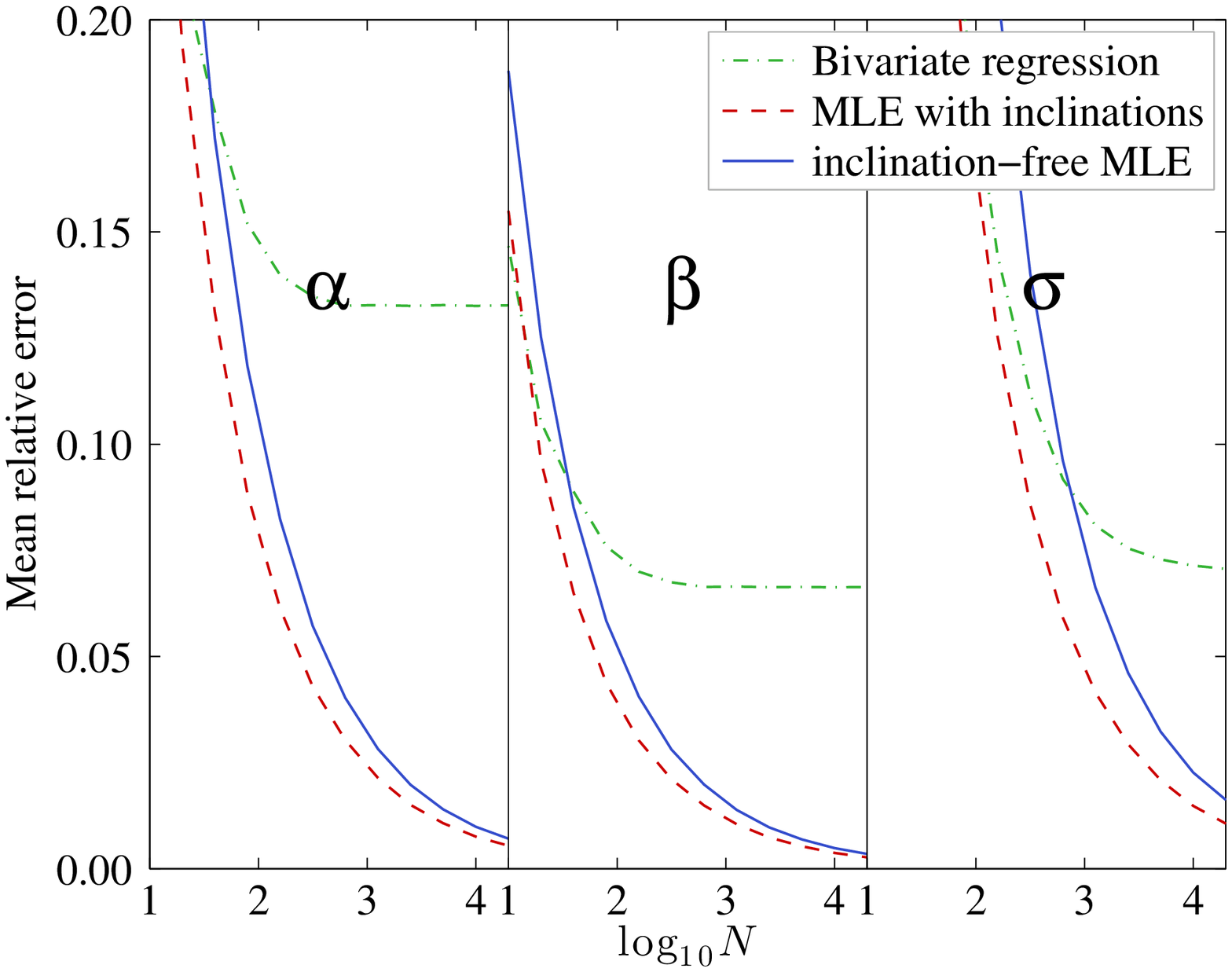}\put(13.5,71.0){\normalsize\textbf{(a)}}\end{overpic} & 
		\begin{overpic}[width=\columnwidth]{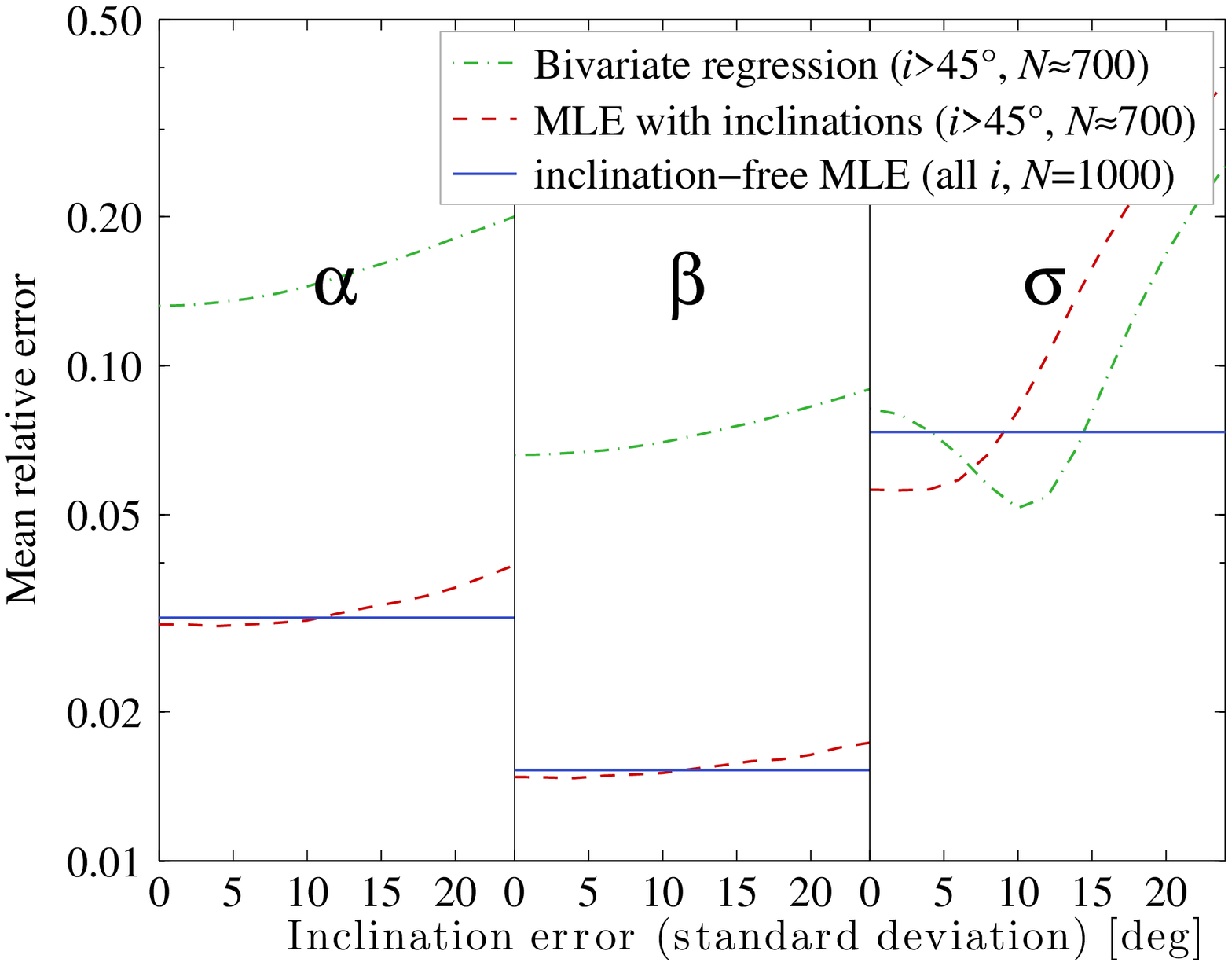}\put(14.5,70.0){\normalsize\textbf{(c)}}\end{overpic} \\
		\begin{overpic}[width=\columnwidth]{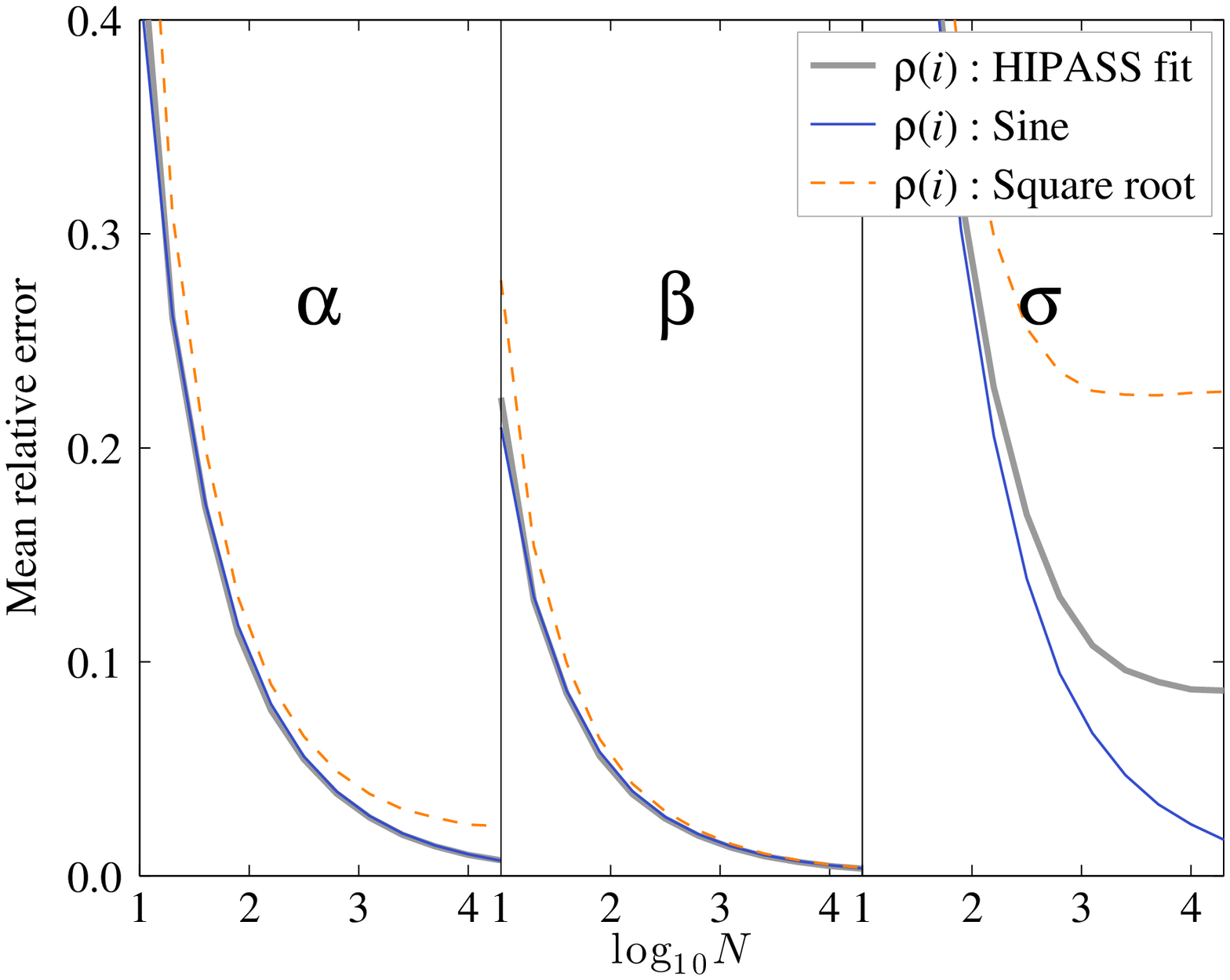}\put(13.5,71.0){\normalsize\textbf{(b)}}\end{overpic} & 
		\begin{overpic}[width=\columnwidth]{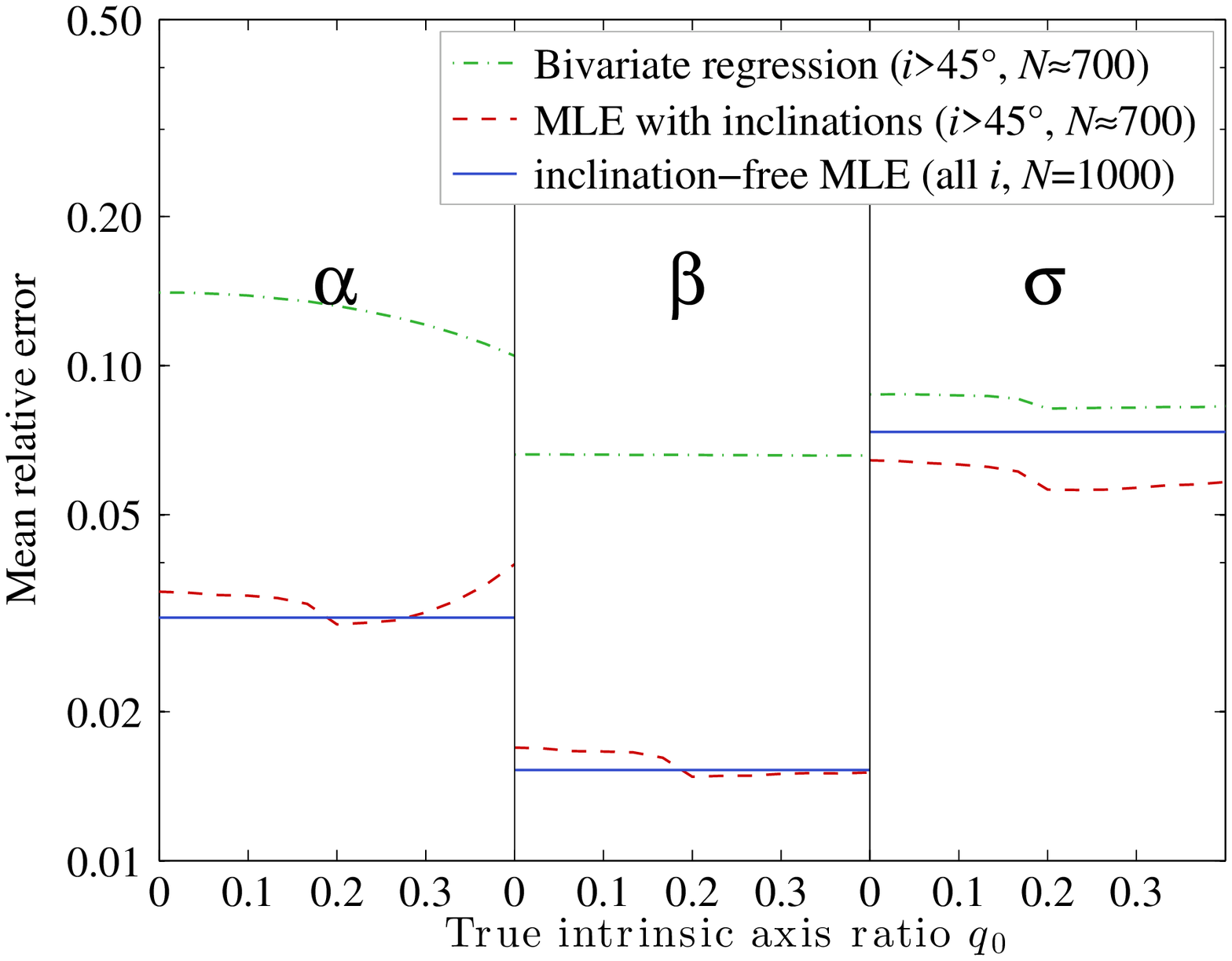}\put(14.5,70.0){\normalsize\textbf{(d)}}\end{overpic} \\
	\end{tabular}
	\caption{Mean relative errors $\ensavg{\Delta\tfr}\equiv(\ensavg{\Delta\alpha},\ensavg{\Delta\beta},\ensavg{\Delta\sigma})$ made in recovering the parameters $\alpha$, $\beta$, and $\sigma$ in the TFR $\logm = \alpha+\beta\logv~(\pm\sigma)$ of different control samples with $N$ galaxies. The four panels are described in detail in different sections: (a) in \ref{subsection_convergence}, (b) in \ref{subsection_inclination_distribution_errors}, (c) and (d) in \ref{subsection_inclination_errors}. Panel (b) uses the same line styles as \fig{inclination_distributions} and refers to the inclination distributions listed in \tab{inclination_distributions}.\col}
	\label{fig_errors}
\end{figure*}

\subsection{Ensemble of control samples}\label{subsection_ensemble}

We can quantify the goodness of a method to estimate the TFR in terms of the error vector $\abs{\tfr-\tfr_0}$, where $\tfr$ denotes the TFR parameters fitted to a control sample with input parameters $\tfr_0$. However, this error depends on the particular random realization of the control sample. Rather than considering a single control sample, we construct an ensemble of $10^3$ random control samples and compute the ensemble mean of the relative error,
\be\label{eq_DT}
	\ensavg{\Delta\tfr}=\left\langle\left|\frac{\tfr-\tfr_0}{\tfr_0}\right|\right\rangle.
\ee
This mean error will be used in \S\ref{subsection_convergence} to \S\ref{subsection_inclination_errors}.

\subsection{Numerical convergence}\label{subsection_convergence}

Let us now \textit{quantify} the performance of the different TFR estimators illustrated in \S\ref{subsection_basic_example} and investigate their asymptotic convergence with the number of galaxies $N$. To this end, we consider 12 \textit{ensembles} of control samples (see \S\ref{subsection_ensemble}) with different $N$, spaced logarithmically between $N=10$ and $N=2\cdot10^4$. The latter corresponds to the largest TFR samples used today \citep[e.g.,][]{Mocz2012}. The mean relative error $\ensavg{\Delta\tfr}$ is plotted against $N$ in \fig{errors}(a) for the different methods.

As expected, $\ensavg{\Delta\tfr}$ decreases monotonically with $N$. However, the error of the bivariate linear regression levels out at a value larger than zero: no matter how many galaxies we use, the linear regression never recovers the input TFR. Formally, this is a direct consequence of the difference between the full likelihood function of \eq{likelihood_fixedi} and the reduced \eq{likelihood_fixedi_simple} used for the bivariate linear regression. Figuratively speaking, the linear regression tends to follow the average slope of the shape spanned in the $(\logw,\logm)$-plane. Depending on the selection function of the data -- here a normal distribution in $\logm$ -- this slope can differ from the actual slope of the input TFR. For mass-limited samples, the slope of the linear regression is likely to be somewhat shallower than the true TFR, but depending on the precise selection criteria, the slope error may also be inverted, or the linear regression may provide perfectly acceptable results. The important message is that the linear regression does not guarantee an optimal recovery of the TFR, although the results might suffice in many cases.

The mean relative errors of the MLEs scale numerically as $\ensavg{\Delta\tfr}\propto N^{-1/2}$ and tend to $0$ as $N\rightarrow\infty$, as expected for Poisson noise. In line with the qualitative conclusion from \fig{basic_example}, the MLE that uses inclinations performs best. However, the relative errors of the inclination-free MLE are only a factor $\sim\!1.5$ larger for any given $N$. Considering that the requirement for optical inclinations tends to reduce number of galaxies $N$ in the parent sample, the inclination-free MLE can typically make use of a larger number $N$ of galaxies than inclination-dependent methods. Hence the relative errors of the inclination-free method become even more competitive.

\subsection{Robustness against inclination distribution errors}\label{subsection_inclination_distribution_errors}

\begin{figure}[b]
	\includegraphics[width=\columnwidth]{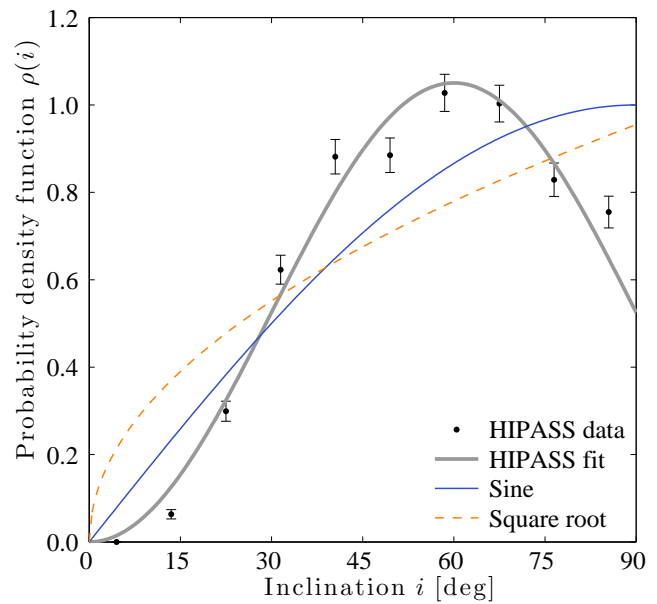}
	\caption{PDFs used for choosing the random inclinations of the galaxy samples considered in \S\ref{subsection_inclination_distribution_errors}. The analytical expressions of these three functions are listed in \tab{inclination_distributions}.}
	\label{fig_inclination_distributions}
\end{figure}

Sections \ref{subsection_basic_example} and \ref{subsection_convergence} demonstrated that the inclination-free MLE can reliably recover the TFR of a galaxy sample that, by construction, has an isotropic distribution of galaxy axes. Expressed alternatively, if the galaxy inclinations satisfy a $\sin i$-distribution, then the inclination-free MLE assuming a $\sin i$-distribution works well. But what if the galaxies in the sample do not obey a $\sin i$-distribution? To investigate this situation, we use control samples with random galaxy inclinations drawn from the modified PDFs shown in \fig{inclination_distributions} and listed in \tab{inclination_distributions}. The `HIPASS-fit' approximates the inclination distribution of the 3,618 tabulated optical counterparts \citep{Doyle2005} of HIPASS detections. These optical inclinations were calculated via \eq{cosi} with $q_0=0.2$. Edge-on galaxies are clearly underrepresented in this distribution, due to the difficulty of detecting the broadest \ha\ lines and due to optical extinction. The `square root' PDF well approximates the predicted inclination distribution of the WALLABY and DINGO HI blind surveys on the Australian SKA Pathfinder (ASKAP) telescope \citep[derived from][]{Duffy2012}.

\begin{table}[b]
	\centering
	\normalsize
	\begin{tabular}{lllc}
	\hline\hline \\ [-1.5ex]
	Name & PDF $\rho(i)$~~~ & Quantile $Q(r)$ & $\ensavg{i}$ \\ [1.0ex]
	\hline \\
	HIPASS fit & $\frac{1-\cos 3i}{\pi/2+1/3}$ & $\approx\frac{\pi}{6}(2r^{0.3}+r^2)$~ & 0.98 \\ [2ex]
	Sine & $\sin i$ & $\arccos r$ & 1.00 \\ [2ex]
	Square root~~ & $\left(\frac{18i}{\pi^3}\right)^{1/2}$ & $\frac{\pi}{2}r^{2/3}$ & 0.94 \\ [2ex]
	\hline
	\hline
	\end{tabular}
	\caption{\upshape\raggedright Three examples of PDFs $\rho(i)$ for the inclinations $i$. In order to draw a random value $i$ from $\rho(i)$, one can draw a uniform random number $r\in[0,1]$ and set $i=Q(r)$, where the quantile function $Q(r)$ is the inverse function of the cumulative distribution function (CDF) $F(i)\equiv\int_0^i d i'\rho(i')$.}
	\label{tab_inclination_distributions}
\end{table}

As in \S\ref{subsection_convergence}, we chose 12 logarithmically spaced values of $N=10,...,2\cdot10^4$. For any $N$ and each of the three PDFs in \fig{inclination_distributions} we construct ensembles of control samples (\S\ref{subsection_ensemble}). Their TFRs are then estimated using the inclination-free MLE, always assuming a $\sin i$-distribution, even if the actual input PDF is different.

The mean relative errors $\ensavg{\Delta\tfr}$ are plotted against $N$ in \fig{errors}(b), with the line styles matching those in \fig{inclination_distributions}. \fig{errors}(b) demonstrates the robustness of the inclination-free MLE against a systematic deviation of the galaxy inclinations from the assumed $\sin i$-distribution. The slope $\beta$ of the TFR remains virtually unaffected, since a systematic deviation of the inclinations affects galaxies of all masses in the same manner and thus merely shifts the TFR horizontally. Intriguingly, even the zero-point $\alpha$ is very robust, the asymptotic mean error being as small as 1-4\% for all PDFs considered here. The main reason for this robustness is the strong dependence of the inclination-free MLE on close-to-edge-on galaxies ($i\gtrsim75^\circ$, thus $w\approx v$): as long as some of these galaxies are present, the precise inclination PDF plays a subordinate role. An additional reason for the robustness of $\alpha$ are the similar means of the three PDFs (last column in \tab{inclination_distributions}). The recovery of intrinsic scatter $\sigma$ is most strongly affected by a systematic deviation of the galaxy inclinations from a $\sin i$-distribution, the mean errors being potentially around 10-20\%. By current observing standards this is nonetheless a small error on the scatter of the TFR.

In conclusion, the inclination-free MLE is robust against a mismatch between the assumed inclination distribution (here a $\sin i$-distribution) and the true inclinations of the galaxy sample. In principle, the true inclination PDF of a galaxy sample could be estimated, for example using a telescope simulator, and then used as the prior PDF in the inclination-free MLE rather than a $\sin i$-distribution. However, given the quantitative results of this section, such extra effort might be unnecessary. 

\subsection{Robustness against inclination errors}\label{subsection_inclination_errors}

An advantage of the inclination-free MLE is its strict independence from inclination measurement errors, which deteriorate the TFRs recovered with inclination-dependent methods. We now quantify the effect of inclination errors based on ensembles of control samples of $N=10^3$ galaxies, where the individual inclinations are perturbed by random and systematic errors.

Random inclination errors are modeled by first drawing the true galaxy inclinations $i_k$ from a $\sin i$-distribution to compute the projected velocities $w_k=v_k\sin i_k$. The true inclinations $i_k$ are then perturbed with Gaussian random noise (mirrored at $i_k=0^\circ$ and $i_k=90^\circ$), resulting in erroneous inclinations $i_k'$ that mimic the observed values. For TFR estimations methods using inclinations, the observed velocities with inclination errors are given by $v_k'\equiv w_k|\sin i_k'|^{-1}$. An inclination cut $i_k'>45^\circ$ is also applied for these methods, since without such an inclination cut their errors $\Delta\tfr$ can become much larger. This cut reduces the number of galaxies available for these inclination-dependent methods to about $N\approx700$. By contrast, the inclination-free MLE can use all $N=10^3$ galaxies, since this method remains strictly unaffected by inclination measurement errors.

\fig{errors}(c) displays the mean relative error $\ensavg{\Delta\tfr}$ as a function of the standard deviation of the inclination errors. This figure reveals that for random inclination errors larger than about $10^\circ$, the TFR is more reliably recovered by the inclination-free MLE. The same result is found for other sample sizes $N$, not shown in the figure. Inclination errors of $10^\circ$ correspond to errors in the axis ratio $q$ of about $0.17, 0.20, 0.16$ at $i=45^\circ, 60^\circ, 75^\circ$. For inclination errors larger than $10^\circ$, it is better not to use any inclination data than to use the uncertain inclinations as exact. Of course, if the uncertainties of the inclinations are well-characterized in terms of probability distributions $\rho_k(i)$, then those distributions can be fed into our MLE to improve on the inclination-free MLE.

Secondly, we address systematic inclination errors arising when the true intrinsic axis ratio $q_0$ deviates uniformly from the assumed intrinsic ratio $q_0'=0.2$. To model this case we use unperturbed galaxy inclinations $i_k$, drawn from a $\sin i$-distribution, to compute the axis ratios $q_k=[q_0^2+(1-q_0^2)\cos^2i_k]^{1/2}$. This equation is then inverted (giving \eq{cosi}), while substituting the true intrinsic axis ratio $q_0$ for $q_0'=0.2$. This results in the erroneous inclinations $i_k'$, which serve to calculate the observed velocities with $q_0$-errors $v_k'=w_k|\sin i_k'|^{-1}$ used in estimating the TFR with the methods that use inclinations. As before, only galaxies with $i_k'>45^\circ$ are used with these methods.

\fig{errors}(d) shows the mean relative error $\ensavg{\Delta\tfr}$ as a function of the true $q_0$ (for fixed assumed $q_0'=0.2$). Although the error $\ensavg{\Delta\tfr}$ slightly increases with  $\abs{q_0-q_0'}$, all methods turn out to be surprisingly robust against errors in the assumed intrinsic axis ratio. With hindsight this finding also justifies the use of a constant, and not well-measured, $q_0$ in many existing TFR studies.

\section{Application to observed data}\label{section_observation}

This section tests the inclination-free MLE to recover the TFRs of real galaxies. We re-analyze a well characterized sample of \ha\ selected galaxies, used by \cite{Meyer2008} to derive a $K$-band and $B$-band TFRs with minimal observational scatter. In selecting galaxies from the HIPASS catalogue for this purpose, \citeauthor{Meyer2008}\ included an inclination cut to nearly edge-on galaxies with $i>75^\circ$ (thus $\sin i\in[0.96,1]$), leading to very reliable circular velocities $v=w/\sin i$. We shall use the same samples, both with and without inclination cut, to compare a standard TFR estimation against the inclination-free MLE. These data allow us to test the inclination-free MLE in more challenging circumstances, including effects such as inclination-dependent extinction that will significantly impact $B$-band magnitudes.

\subsection{Reference Tully-Fisher sample}\label{subsection_empirical_data}

The study of the $K$-band and $B$-band TFRs by \cite{Meyer2008} relies on four data sets: the \ha\ Parkes All-Sky Survey (HIPASS) Catalogue (HICAT, \citealp{Meyer2004,Zwaan2004}) which provides the initial \ha\ selected sample with \ha\ linewidths; the HIPASS Optical Catalogue (HOPCAT, \citealp{Doyle2005}), which provides more accurate galaxy positions and aspect ratios $q$  used to calculate the inclinations $i$ via \eq{cosi} with $q_0=0.12$; the Two Micron All-Sky Survey (2MASS, \citealp{Jarrett2000}) providing the $M_K$ magnitudes; and the ESO Lauberts \& Valentijn catalogue (ESO-LV, \citealp{Lauberts1989}) providing the $M_B$ magnitudes. Those magnitudes have been corrected for redshift ($k$-correction), extinction by dust in the Milky Way, and intrinsic extinction by the source (details in \S3.1 of \citealp{Meyer2008}). The linewidths are taken as the \ha\ linewidths $W_{50}$, measured at the 50\% level of the peak flux density and corrected for relativistic broadening, instrumental broadening, and turbulent motion (\S3.2 of \citeauthor{Meyer2008}). To study the $K$-band and $B$-band TFRs, the log-mass $\logm$ of the generic TFR (\eq{observable_tfr}) is substituted for $M_K$ and $M_B$, and the projected velocities $w$ are taken as $w=W_{50}/2$.

Departing from the full sample of galaxies with sufficient data in HICAT, HOPCAT, 2MASS, and ESO-LV, \cite{Meyer2008} derived raw $K$-band $B$-band TFRs (see their Figure 3). They discovered that most of the scatter in these TFRs was observational, being correlated to observer-dependent quantities such as inclination, apparent size, and large-scale flows. Restricting the sample to inclinations $i>75^\circ$, effective semi-major axes $r_{max}>15''$, velocities $cz>1000~\kms$, and positions less than $20^\circ$ from the dipole equator of the cosmic microwave background (CMB), led to clean subsamples with tight $K$-band and $B$-band TFRs over six magnitudes; the bivariate regressions quoted by \citeauthor{Meyer2008} are $\beta_K=-9.38\pm0.19$, $\sigma_K=0.25\rm$, $\beta_B=-8.51$, $\sigma_B=0.33$.

\subsection{Recovery of the Tully-Fisher Relation}\label{subsection_KTFR}

\begin{figure*}[t]
	\centering
	\begin{tabular}{cc}
		\begin{overpic}[width=\columnwidth]{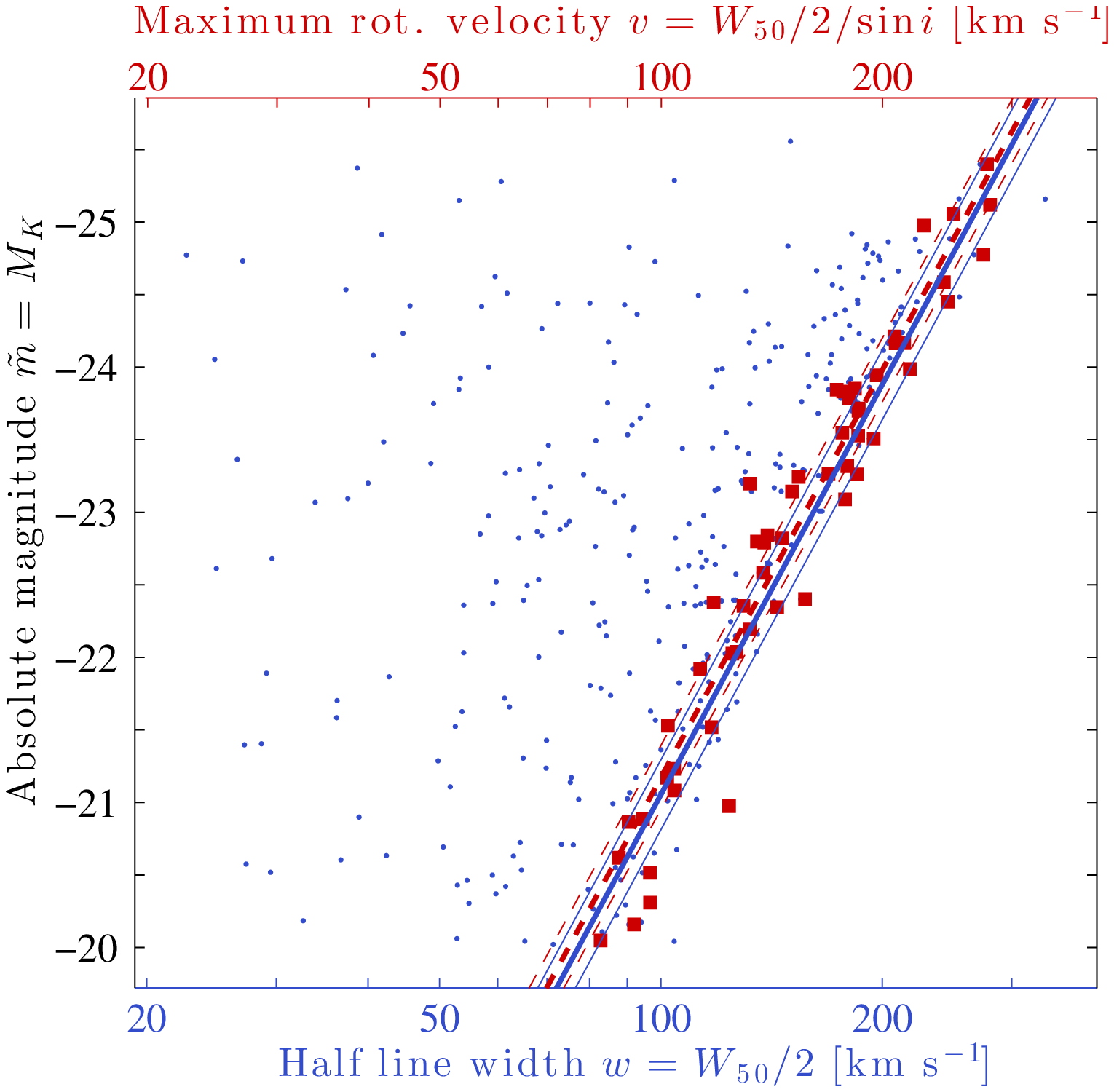}\put(13,70.0){\normalsize\textbf{~}}\end{overpic} & 
		\begin{overpic}[width=\columnwidth]{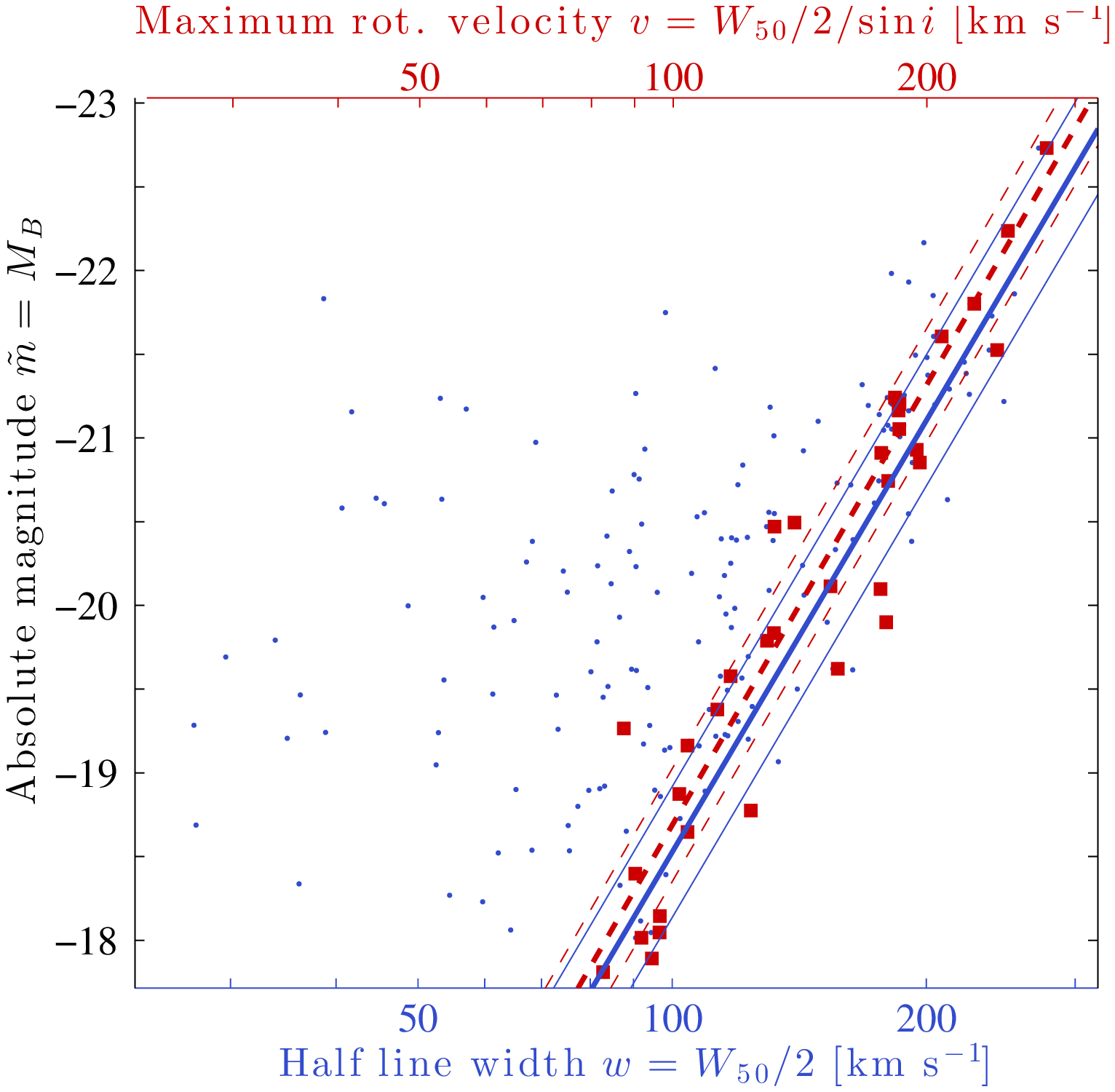}\put(14,70.0){\normalsize\textbf{~}}\end{overpic} \\
	\end{tabular}
	\caption{Absolute $K$-band (left) and $B$-band (right) magnitudes, plotted against $v$ (red big dots, $i>75^\circ$) and $w$ (blue small dots, all $i$). The lines represent the fitted TFRs with 1-$\sigma$ scatter. Red dashed lines show the TFRs obtained by applying the MLE using inclination measurements, i.e, these lines are the most likely fits to the red big dots. In turn, the blue solid lines are obtained without inclination data, by applying the inclination-free MLE, which assumes an isotropic prior on the inclinations, i.e., $\rho(i)=\sin i$. In other words, these fits are based purely on the scattered blue small dots. The numerical values of all TFRs are listed in \tab{real_tfr}.\col}
	\label{fig_real_tfr}
\end{figure*}

The aforementioned clean subsamples of nearly edge-on ($i>75^\circ$) galaxies constructed by \cite{Meyer2008} contain 55 ($K$-band) and 36 ($B$-band) objects. They are displayed in \fig{real_tfr} in the $(\logv,\logm)$-plane (big dots). The red dashed lines represent the corresponding TFRs, fitted directly to the $(\logv,\logm)$-points using the MLE that simultaneously optimizes all three parameters $\tfr=(\alpha,\beta,\sigma)$ to minimize \eq{likelihood_fixedi}. The fitted TFR parameters (see \tab{real_tfr}) are consistent with those resulting from the bivariate regression applied by \citeauthor{Meyer2008}, although regressions are theoretically less exact (see \S\ref{subsection_MLE_with_inclinations}).

\begin{table}[t]
	\centering
	\normalsize
	\begin{tabular}{lcccc}
	\hline\hline \\ [-1.5ex]
	Sample & $N$ & Offset $\alpha$ & Slope $\beta$ & Scatter $\sigma$ \\[1.0ex]
	\hline \\[-1.0ex]
$K$, $i>75^\circ$ & 55 & $+2.5\pm 0.7$ & $-9.3\pm 0.3$ & $0.22\pm0.06$ \\ 
$K$, all $i$ & 351 & $+2.3\pm 0.6$ & $-9.4\pm 0.3$ & $0.24\pm0.06$ \\ 
$B$, $i>75^\circ$ & 36 & $+1.2\pm 0.8$ & $-8.7\pm 0.3$ & $0.34\pm0.06$ \\ 
$B$, all $i$ & 180 & $+1.4\pm 0.6$ & $-8.6\pm 0.3$ & $0.39\pm0.04$ \\ 
	[1.0ex]\hline\hline
	\end{tabular}
	\caption{\upshape\raggedright Numerical values for the TFRs shown in \fig{real_tfr}. The TFRs of the edge-on samples with $i>75^\circ$ were obtained by applying the MLE using full information on inclinations, i.e., inclinations were used both to calculate the velocities $v$ from the linewidths and to correct to the magnitudes for internal extinction. By contrast, the TFRs fitted to the full samples (all $i$) use strictly no inclinations. These TFRs have been obtained by applying the inclination-free MLE that assume an isotropic distribution of the galaxies, i.e., $\rho(i)=\sin i$. Parameters are given with $68\%$ confidence intervals obtained by Bootstrapping the data.}
	\label{tab_real_tfr}
\end{table}

To check the inclination-free MLE, we now drop the selection criterion $i>75^\circ$ and include galaxies of all inclinations. In practice no inclination data often means no resolved images, thus no values for the semi-major axes, and hence no possibility for an explicit size selection. We therefore couple the removal of the inclination cut with the removal of the size cut and instead add an absolute magnitude cut $M_K<-20$ and $M_B<-18$ to still suppress the gas-dominated dwarf galaxies that were excluded by the size cut ($r_{max}>15''$). This results in samples of 351 ($K$-band) and 180 ($B$-band) galaxies of all inclinations. \fig{real_tfr} shows this sample in the $(\logw,\logm)$-plane (small dots). Further, we also remove the inclination-dependent extinction correction from the data (by suppressing the term $A_i$ in Equation (1) of \citealp{Meyer2008}). These corrections have a significant (0.2~mag average) effect for the $B$-band magnitudes, but are negligible for $K$-band magnitudes.

To recover the TFRs we then feed the absolute magnitudes $\logm_k$ (without corrections for internal extinction) and the observed projected velocites $\logw_k$ into the MLE, strictly using no information on inclination. The likelihood function of \eq{likelihood} is maximized assuming random galaxy inclinations ($\rho(i)=\sin i$) and accounting for the unknown inclination-dependent extinction (see \S\ref{subsection_real_data}) $\logm_k(i)=\logm_k+A_i$ with the extinction factor $A_i$ defined as in \citealp{Meyer2008}. This results in the TFRs plotted as blue solid lines in \fig{real_tfr}.

\fig{real_tfr} demonstrates that the inclination-free MLE provides similar TFRs than the observationally more expensive approach, which uses inclinations. To quantify this similarity, both methods have been re-sampled $10^3$ times with random half-sized subsamples. This bootstrapping provides the 68\% confidence intervals listed in \tab{real_tfr}. All three TFR parameters $\alpha$, $\beta$, and $\sigma$ of the two methods are consistent in that they deviate by less than one standard deviation. Given the high co-variance of these three parameters, the same conclusion applies to the full vector $\tfr=(\alpha,\beta,\sigma)$. In summary, this illustration based on real data suggests that the inclination-free MLE can recover the TFR within errors comparable to the typical shot noise uncertainties.

\section{Conclusions}\label{section_conclusion}

This work introduced a mathematical framework for studying the Tully-Fisher relation (TFR) based on maximum likelihood estimation (MLE). This framework permits the estimation of the TFR in galaxy samples with uncertain inclinations, described by a probability distribution $\rho_k(i)$ for every galaxy $k$. This method consists of maximizing the likelihood function of \eq{likelihood}, where $\rho_k(\tfr)$ is given in \eq{main_rho2} for Gaussian measurement uncertainties. This method also applies to the extreme situation, where no inclinations are known at all. In this case, the inclination priors can be replaced by an isotropic distribution, $\rho_k(i)=\sin i$ (unless a better prior is available). Our analysis of this `inclination-free MLE' has led to the following results:
\begin{itemize}
	\item \textit{No galaxy inclinations are needed to recover the TFR.} If a galaxy sample obeys a TFR with normal scatter, the inclination-free MLE generally recovers the three TFR parameters (zero-point, slope, scatter) with statistical errors about 1.5-times larger than the best estimates based on perfectly known galaxy inclinations with zero uncertainty. 
	\item \textit{The inclination-free MLE is robust.} Its independence of galaxy inclinations makes it independent of inclination errors, be they measurement errors, model errors of $q_0$, or real misalignments between the disk used for linewidth measurements (e.g., \ha~disk) and the disk used for inclination measurements (e.g., optical disk). In a realistic scenario, where inclination errors are larger than $10^\circ$ and a typical inclination selection of $i>45^\circ$ must be applied for inclination-dependent methods, the inclination-free MLE performs in fact better than methods that use the measured inclinations while treating them as exact. Even if the assumed PDF $\rho(i)$ (typically a $\sin i$-distribution) of the galaxy inclinations is inaccurate, the errors of the TFR parameters remain small (\fig{errors}(b)).
	\item \textit{The inclination-free MLE is flexible.} If desired, this method can account for a PDF $\rho(i)$ that differs from a $\sin i$-distribution. The method can also be extended in a straightforward way to deal with detection limits, complex measurement uncertainties, erroneous outliers in the data, elliptical/irregular galaxies, and inclination-dependent extinction.
\end{itemize}

The inclination-free MLE is potentially crucial for future surveys, namely \ha\ blind surveys with the Australian SKA Pathfinder (ASKAP), the South African SKA Pathfinder (MeerKAT), and the SKA. In these large surveys, ancillary optical data may not have sufficient spatial resolution for reliable inclination measurements. In this case, the inclination-free MLE can recover various TFRs, including the stellar mass and baryon mass TFRs. By removing the need for well-resolved spatial imaging, the inclination-free MLE also enables the TFR to be recovered at higher redshifts than might otherwise be possible, offering a new method of probing the cosmic evolution of the TFR. Should optical counterparts be completely unavailable, one might still use the inclination-free MLE to derive a \ha~mass TFR, as done for the HIPASS data in Figure 13a of \cite{Obreschkow2009b}.

Indirectly, the inclination-free MLE can be used to constrain the inclinations of galaxies with otherwise unknown inclinations. Once the TFR parameters $\tfr$ have been determined using the inclination-free MLE, every galaxy $k$ has a well-defined PDF $\rho_k(\logv)$ for its circular velocity (\eq{rho_tfr}). Using $v_k=w_k/\sin i_k$, this PDF can be rewritten as a PDF for the inclinations $i_k$. When multiplied by the \textit{prior} PDF $\rho(i)=\sin i$, this results in a \textit{posterior} PDF $\rho^{post}_k(i)$. This posterior PDF\footnote{Explicitly, $\rho^{post}_k(i)=\rho(i)\exp[-[\log_{10}(\sin(i)/r)]^2/(2\sigma_{tot,k}'^2)]$, where $\sigma_{tot,k}'\equiv\sigma_{tot,k}/\beta$ is the horizontal scatter and $r\equiv w_k/\avg{v}_k$ with $\avg{v}_k=10^{(\logm_k-\alpha)/\beta}$ being the mean circular velocity predicted by the TFR. The width of $\rho^{post}_k(i)$ increases monotonically with $\sigma_{tot,k}'$, and as $\sigma_{tot,k}\rightarrow\infty$ we have $\rho^{post}_k(i)=\rho(i)$. This simply means that no constraints on inclinations could be found via the TFR, if the scatter of the latter were too large.} could be used, for instance, to study correlations between inclinations and other galaxy properties -- an idea to be explored in forthcoming studies.

This project was supported by the UWA Research Collaboration Award 2013 (PG12104401). We thank the anonymous referee for significant contributions.


\appendix

\section{A. Advanced extensions of the MLE approach}\label{appendix_extensions}

This section sketches out possible extensions that allow the MLE of Section \ref{section_method} to be applied to more advanced situations. In particular, we consider (i) complex detection limits affecting the linewidth selection, (ii) samples contaminated by merging and elliptical galaxies, and (iii) non-linear TFRs.

\subsection{A.1.~Complex detection limits}\label{subsection_sample_truncation}

The derivation of the MLE in \S\ref{subsection_generic_mle} assumed that a galaxy can be detected with a probability independent of $w$. In reality this is an approximation, since large linewidths contain more noise and linewidths smaller than the channel width might be misinterpreted as radio frequency interferences (RFIs) or other artifacts. If a line is detected with a non-uniform detection rate $f(\logw|\logm)\in[0,1]$, then $\rho_k(\logm,\logw|\tfr)$ in \eq{rhok_general} must be substituted for
\be\label{eq_rhoast}
	\rho^\ast_k(\logm,\logw|\tfr) \equiv \frac{f(\logw|\logm)\rho_k(\logm,\logw|\tfr)}{\int d\logw\,f(\logw|\logm)\rho_k(\logm,\logw|\tfr)}.
\ee

As an example, let us consider the case where galaxies are detected perfectly if $\logw\geq\logw_0$ and undetected otherwise. Formally, $f(\logw|\logm)=1$, if $w\geq w_0$, and $f(\logw|\logm)=0$ otherwise. In this case, \eq{rhoast} becomes
\be\label{eq_rhostar_example}
	\rho^\ast_k(\logm,\logw|\tfr) = \begin{cases}
		S_k^{-1}(\logm,\tfr)\rho_k(\logm,\logw|\tfr), & \!\!\!\!\mbox{if } w\geq w_0, \\
		0, & \!\!\!\!\mbox{if } w<w_0,
	\end{cases}
\ee
where $S_k(\logm,\tfr)\equiv\int_{\logw_0}^{\infty}d\logw\,\rho_k(\logm,\logw|\tfr)$ can be integrated partially to
\be
	S_k(\logm,\tfr) = \frac{1}{2}\int_0^{\pi/2}\!\!\!\!\!\!d i\,\rho_k(i)\left[\erf\Big(\tfrac{\logm-\alpha-\beta\logw_0+\beta\log_{10}\sin i}{\sqrt{2}\sigma}\Big)+1\right].
\ee
Substituting $\rho_k(\logm,\logw|\tfr)$ for $\rho^\ast_k(\logm,\logw|\tfr)$ in \eq{rhok_general} and assuming Gaussian measurement uncertainties for $\logw$ and $\logm$ (as in \S\ref{subsection_generic_mle}), \eq{rhok_general} solves in good approximated to
\be\label{eq_new_rhok}
	\rho_k(\tfr) = \frac{\frac{\beta}{\sqrt{2\pi}\sigma_{tot,k}}\int_0^{\pi/2}d i\,\rho_k(i) \exp\Big[-\tfrac{(\logm_k-\alpha-\beta\logw_k+\beta\log_{10}\sin i)^2}{2\sigma_{tot,k}^2}\Big]}
	{\frac{1}{2}\int_0^{\pi/2}d i\,\rho_k(i)\left[\erf\Big(\tfrac{\logm-\alpha-\beta\logw_0+\beta\log_{10}\sin i}{\sqrt{2}\sigma_{tot,k}}\Big)+1\right]}\,.
\ee

\begin{figure}[h]
	\centering
	\includegraphics[width=8.5cm]{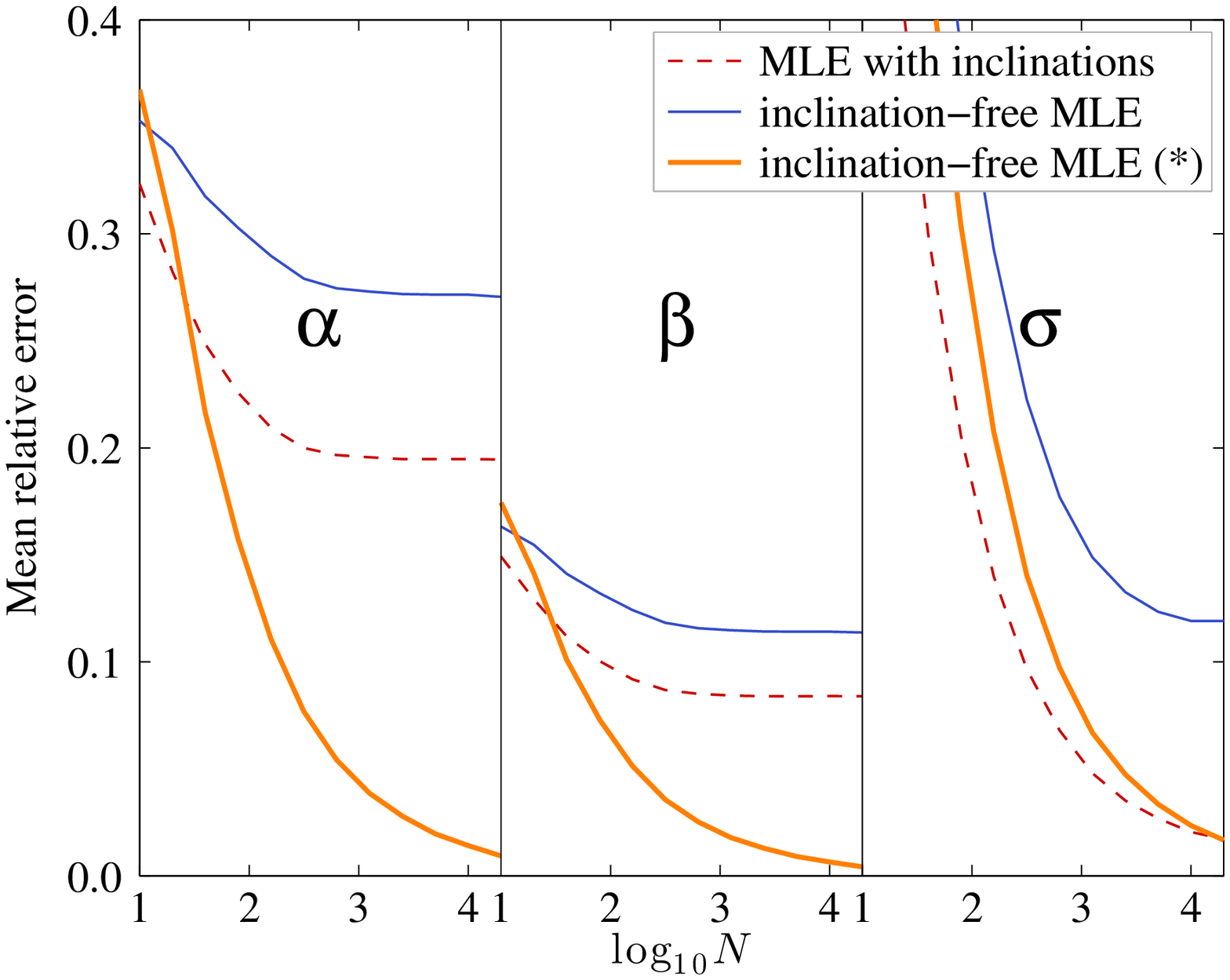}
	\caption{Mean relative errors $\ensavg{\Delta\tfr}\equiv(\ensavg{\Delta\alpha},\ensavg{\Delta\beta},\ensavg{\Delta\sigma})$ made when estimating the TFR of a simulated control sample as described in \S\ref{subsection_control_sample}, but truncated to $w>w_0=1.5$. The three line types represent different methods to recover the TFR. Two of those methods are identical to those in \fig{errors}(a), while the last one (yellow thick solid line) is the inclination-free MLE that explicitly accounts for the $w$-truncation, using the formalism of \S\ref{subsection_sample_truncation}.\col}
	\label{fig_convergence_truncated}
\end{figure}

In summary, using \eq{new_rhok} instead of \eq{main_rho2} when maximizing the likelihood function of \eq{likelihood} allows an optimal recovery of the TFR in a sample truncated to $\logw\geq \logw_0$. As an illustration we repeat the convergence study of \S\ref{subsection_convergence}, while removing all galaxies with $\logw<\logw_0=1.5$ from the control samples. \fig{convergence_truncated} shows that the errors $\ensavg{\Delta\tfr}$ that asymptotically vanished in \fig{errors}(a), no longer vanish now. However, using \eq{new_rhok} instead of \eq{main_rho2}, the TFR is again recovered with high precision, as shown by the yellow thick solid lines in \fig{convergence_truncated}.

\subsection{A.2.~Contamination by merging and elliptical galaxies}

The inclination-free MLE can be applied to galaxy samples with no spatially resolved imaging. In that case, the morphology-based removal of merging systems and elliptical galaxies is impossible. This raises the concern of a contamination by merging and elliptical galaxies, which do not obey the simple TFR of spiral galaxies \citep{Faber1976,Kassin2007}. For most studies this is a negligible concern, because merging systems are rare in the local universe and elliptical galaxies can be approximately excluded based on color. Moreover, in \ha-selected samples, such as HIPASS, elliptical galaxies are rarely detected because of their small \ha~fractions and broad linewidhts (\S3.3 in \citealp{Obreschkow2013b}).

In some cases, one might deliberately include merging and elliptical galaxies to study a generalized TFR, where the rotation velocity $v$ is substituted for the kinematic estimator $S_{0.5}\equiv\sqrt{0.5v^2+s^2}$ with $s$ being the baryon velocity dispersion. According to \cite{Weiner2006} (\S 5.2 therein) $S_{0.5}$ is a better tracer of the total velocity dispersion of the galaxy-halo system than $v$. \cite{Kassin2007} have indeed shown that spiral, elliptical, and merging galaxies all fall on the same $S_{0.5}$-TFR in terms of slope and zero-point. Following the philosophy of this work, one could use an inclination-free MLE approach to estimate the $S_{0.5}$-TFR. Formally this requires substituting $v$ for $S_{0.5}\equiv\sqrt{0.5v^2+s^2}=\sqrt{0.5(w/\!\sin i)^2+s^2}$. In logarithmic form, $\tilde{S}_{0.5}=0.5\log_{10}(0.5(w/\!\sin i)^2+s^2)=0.5\log_{10}[(x/\!\sin i)^2+1]+\logs$, where $x\equiv w/(\sqrt{2}\,s)$. Following an analogous development to \S\ref{subsection_generic_mle}, the most likely $S_{0.5}$-TFR is then given by maximizing \eq{likelihood}, where
\be\label{eq_main_rho_kassin}
	\rho_k(\tfr) =\,\frac{\beta}{\sqrt{2\pi}\sigma_{tot,k}}\int_0^{\pi/2}\!\!\d i\,\rho_k(i)\exp\Big[-\tfrac{(\logm_k-\alpha-0.5\beta\log_{10}[(x_k/\!\sin i)^2+1]-\beta\logs_k}{2\sigma_{tot,k}^2}\Big],
\ee
with $x_k=10^{\logw_k-\logs_k}/\sqrt{2}$.

\subsection{A.3.~Verification of the TFR hypothesis}\label{subsection_tfr_deviations}

The MLE approach relies on the hypothesis that the data obeys a linear relation in the $(\logv,\logm)$-plane with Gaussian scatter. In reality, the Gaussianity of the scatter is consistent with the state-of-the-art baryonic TFR \citep[\S2.6.4 in ][]{McGaugh2012}, but the linearity of the TFR is less verified in nature. For example, the slope of the stellar mass TFR is found to increase at small rotation velocities ($v\lesssim100~\kms$), where the contribution of gas to the baryon budget becomes significant \citep{McGaugh2000}. If the assumption of a linear TFR with Gaussian scatter fails, the MLE approach introduced in this paper does not optimally recover the TFR -- neither do linear regression techniques.

How can we verify whether the null hypothesis of a linear TFR with Gaussian scatter applies? If all galaxies have known inclinations and thus known circular velocities, a reduced $\chi^2$ test can be applied, where $\chi^2_{red}=\sum_{k=1}^Ng_k(\Delta\logv_k)^2/(\sigma_{tot,k}/\beta)^2$ with $\Delta\logv_k=\logv_k-(\logm_k-\alpha)/\beta$. If and only if $\chi^2_{red}\lesssim1$, the null hypothesis is consistent with the data. The situation is more subtle if inclinations are unavailable and the TFR is found via the inclination-free MLE. In that case, the estimated TFR can be used to assign a PDF for the circular velocities $\rho_k(\logv)$ to every galaxy $k$. Those can be converted into a posterior PDF for the measured half-linewidth $\rho_k(\logw)$ (\eq{main_rho1}). In the Gaussian approximation, one can then use $\chi^2_{red}=\sum_{k=1}^Ng_k\,(\logw_k-\ensavg{\logw_k})^2/\ensavg{(\logw_k-\ensavg{\logw_k})^2}$, where $\ensavg{\logw_k}\equiv\int d\logw\,\logw\,\rho_k(\logw)$ is the expectation value of $\logw_k$ given the TFR and $\ensavg{(\logw_k-\ensavg{\logw_k})^2}\equiv\int d\logw\,(\logw-\ensavg{\logw_k})^2\,\rho_k(\logw)$ is the variance. If the null hypothesis fails, that is $\chi^2_{red}\gg1$, a linear TFR with Gaussian scatter does not sufficiently explain the data. The model can then be extended by modifying \eq{rho_tfr}, which defines the linearity of the TFR, and/or \eq{def_tfr}, which defines the Gaussianity of the scatter. The other equations in Section \ref{section_method} must be modified accordingly.


\end{document}